\documentclass[5p,twocolumn]{elsarticle}

\usepackage{graphicx}
\usepackage{dcolumn}
\usepackage{pifont}
\usepackage{bm}
\usepackage{multirow}
\usepackage{amsmath}
\usepackage{float}
\usepackage{txfonts}
\usepackage{textcomp}
\usepackage[version=3]{mhchem}

\def\mathbi#1{\textbf{\em #1}}

\usepackage[usenames,dvipsnames]{xcolor}
\definecolor{cream}{RGB}{222,217,201}
\definecolor{red}{RGB}{225,0,0}

\journal{arXiv}

\begin{document}
\title{Molecular dynamics study of the effect of moisture and porosity on thermal conductivity of tobermorite 14 \AA}

\author[kimuniv-m]{Song-Nam Hong}
\author[kimuniv-m]{Chol-Jun Yu\corref{cor}}
\cortext[cor]{Corresponding author}
\ead{cj.yu@ryongnamsan.edu.kp}
\author[kimuniv-m]{Kum-Chol Ri}
\author[kimuniv-m]{Ju-Myong Han}
\author[kumsu]{Byong-Hyok Ri}

\address[kimuniv-m]{Chair of Computational Materials Design (CMD), Faculty of Materials Science, Kim Il Sung University, Ryongnam-Dong, Taesong District, Pyongyang, Democratic People's Republic of Korea}
\address[kumsu]{Changhun-Kumsu Agency for New Technology Exchange, Mangyongdae District, Pyongyang, Democratic People's Republic of Korea}

\begin{abstract}
The effect of moisture and porosity on thermal conductivity of tobermorite 14 \AA~as the major component of cement paste is studied by using molecular dynamics simulation with ClayFF potential.
The calculated results show that the thermal conductivity increases monotonically as the moisture content by mass within the interior pores increases and the slope of the linear fitting function decreases as the porosity increases.
Meanwhile, the normalized thermal conductivity is found to increase exponentially as increasing the moisture content by volume.
Phonon density of states of porous and moist tobermorite 14 \AA~is used to explain the contribution of individual atoms and molecules to the thermal properties.
The results can be potentially used to design higher thermal insulating materials with cement and concrete for energy saving buildings.
\end{abstract}


\maketitle

\section{\label{sec:intro}Introduction}
In recent years, there has been a rapid increase in concerns for depletion of fossil fuel resources, green house gas emission and steady increase of energy demand~\cite{Chu}.
To resolve such huge and complex problems, we should surely raise the volume of energy production as much as possible, using sustainable and clean energy sources such as solar, wind and tidal power, but the efficient consumption of produced energy is also very important.
In particular, it is vital to save the building energy for its construction and operation, because buildings are recognized to be responsible for about 40\% of global energy use and almost 50\% of world greenhouse gas emission~\cite{Rashid15rser}.
Moreover, due to much higher energy portion ($\sim$80\%) of building operation (heating, cooling, and ventilation) than its construction~\cite{Martinez}, cement concrete and paste composing the building wall should have low thermal conductivity.

So far, many research works have been reported for the thermal conductivity of cementitious materials~\cite{Yoon,Xu,Bentz,Qomi1,QingJin,Wu,Hansen,Mounanga}.
Through the works, the main factors influencing the thermal conductivity of cementitious materials have been identified as composition, porosity, moisture content, water-cement (w/c) ratio and curing age.
Among them, the moisture content has been accepted to be one of the most critical factors on the thermal conductivity of cement paste and concrete.
In addition, durability, service span and ecological environment of concrete buildings are also mostly related to the moisture content of concrete.
Although a lot of macroscopic measurements have been performed to reveal the effect of moisture~\cite{QingJin,Santos,Campanale,Belkharchouche,Gomes}, a microscopic understanding of thermal conductivity of cement paste is scarcely provided.

It is generally accepted that in cement paste water exists in the interior pores, including interlayer space with a size of $\sim$1 nm, gel pores (1$\sim$5 nm) and capillary pores (5$\sim$50 nm)~\cite{Mehta2014}.
Here the gel pores are spaces between the nano-grains, while the capillary pores are empty spaces created during the hydration of cement.
These pores gradually decrease with age.
When cement reacts with water, some hydration products are generated, including C-S-H (CaO$-$\ce{SiO2}$-$\ce{H2O}) gel, CH (portlandite), and hydrogarnet.
Among them, the C-S-H gel has the largest volume, and therefore, it is responsible for the properties of cement paste and concrete.
Although the structure of C-S-H gel is yet explicitly unrevealed, previous experimental works performed using the high-resolution transmission electron microscopy (TEM) and extensive X-ray diffraction (XRD) indicated that the C-S-H gel mimics the atomistic structure of tobermorite family and jennite~\cite{Bonaccorsi-tober14,Bonaccorsi-jen}.
In many previous {\it ab initio} and molecular dynamics simulations, the structural, mechanical and thermodynamical properties of cement paste have been investigated based on tobermorite and jennite models.

The tobermorite family includes three different polymorphs such as tobermorite 9 \AA, tobermorite 11 \AA~and tobermorite 14 \AA~according to the interlayer distance~\cite{Bonaccorsi-tober14,Merlino99,Merlino01}.
Previous experimental results revealed that tobermorite 14 \AA~ may transform to tobermorite 11 \AA~at 80-100 \textcelsius, and subsequently to tobermorite 9 \AA~at 300 \textcelsius~\cite{Merlino99,Yu,Biagioni}.
Molecular dynamics (MD) simulations have been performed to explain the various properties of cement paste, mostly for microstructural and mechanical properties~\cite{Manzano2009,Al-Ostaz,Fan,Hajilar,Mutisya}.
Some studies have been performed on the transport properties of confined water in cement paste~\cite{Dongshuai,Hou1,Hou2,Bonnaud,Zhang2017}.
To the best of our literature search, however, studies for revealing the thermal conductivity of tobermorite 14 \AA~at molecular scale have not yet been provided.

In this work, we investigate the influence of porosity and moisture content on the thermal conductivity of tobermorite 14 \AA~to mimic cement paste by using MD simulations.
It is worth noting that tobermorite 14 \AA~is the best analogous to the C-S-H gel~\cite{Taylor1986}.
In Section~\ref{sec_method}, computational methods are described, including the force field potential and tobermorite 14 \AA~models with various porosities and moisture contents.
Section~\ref{sec_result} shows the simulation results in comparison with the previous experimental and simulation works.
Finally, conclusions of this paper are provided in Section~\ref{sec_con}.

\section{\label{sec_method}Computational methods}
\subsection{\label{sub_model}Structural models for porous and moist tobermorite 14 \AA}
\begin{figure}[!b]
\centering
\includegraphics[clip=true,scale=0.15]{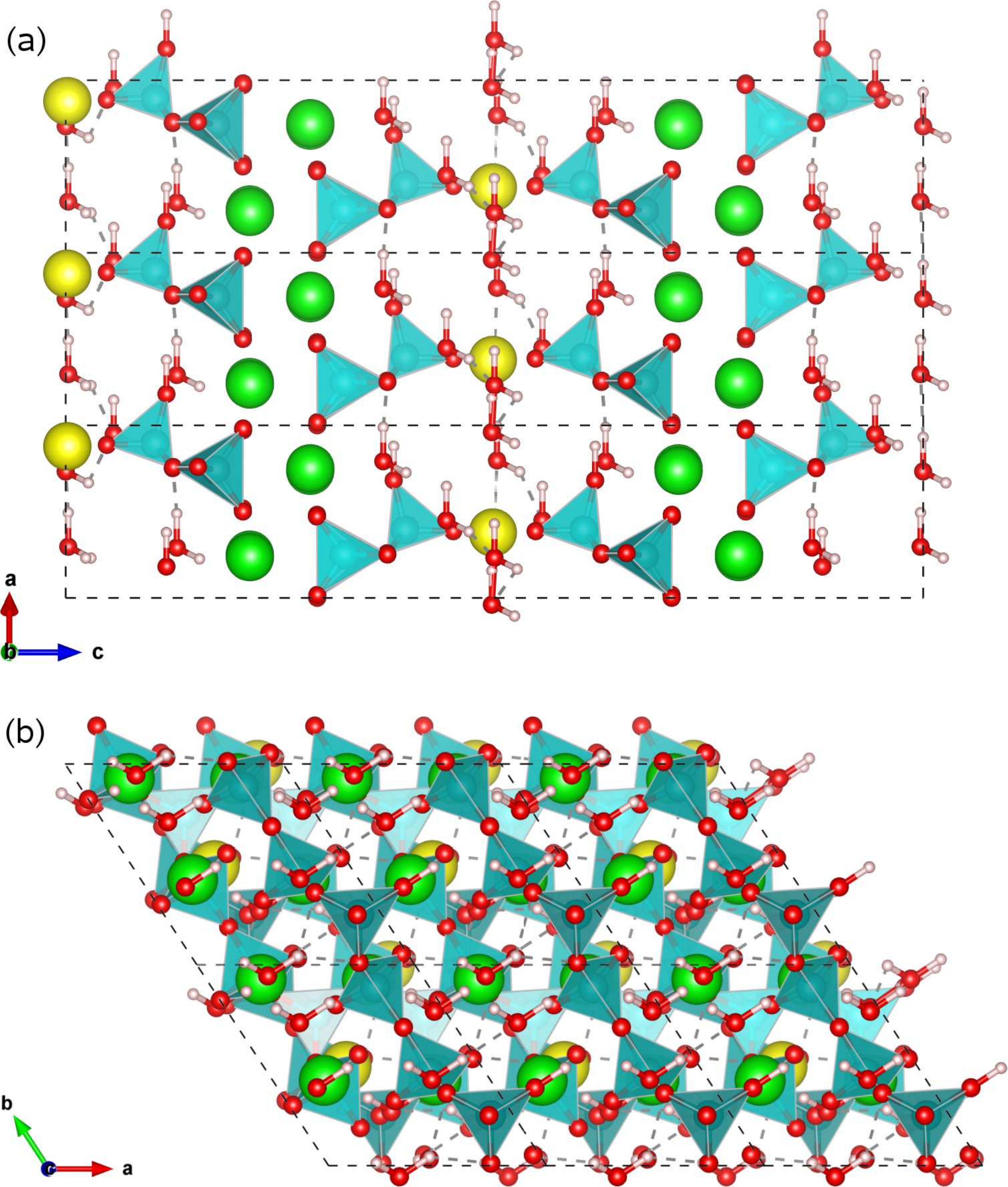}
\caption{Polyhedral view of atomistic structure of tobermorite 14 \AA~unit cell in (a) (010) and (b) (001) planes.
Yellow, green, cyan, red and white balls represent the interlayer Ca, intralayer Ca, Si, O and H atoms respectively.}
\label{fig1}
\end{figure}
The structural model of tobermorite 14 \AA~was based on the ordered layer structure of monoclinic phase with a space group of {\it B11b} and the chemical formula of \ce{Ca5Si6O_{16}(OH)2\cdot}7\ce{H2O}~\cite{Bonaccorsi-tober14}.
Although calcium atoms and water molecules in interlayer space can occupy the sites randomly in natural sample of tobermorite 14 \AA, they alternately reside in the locations in the case of such ordered model.
Figure~\ref{fig1} shows the unit cell of tobermorite 14 \AA.

\begin{figure}[!t]
\centering
\includegraphics[clip=true,scale=0.15]{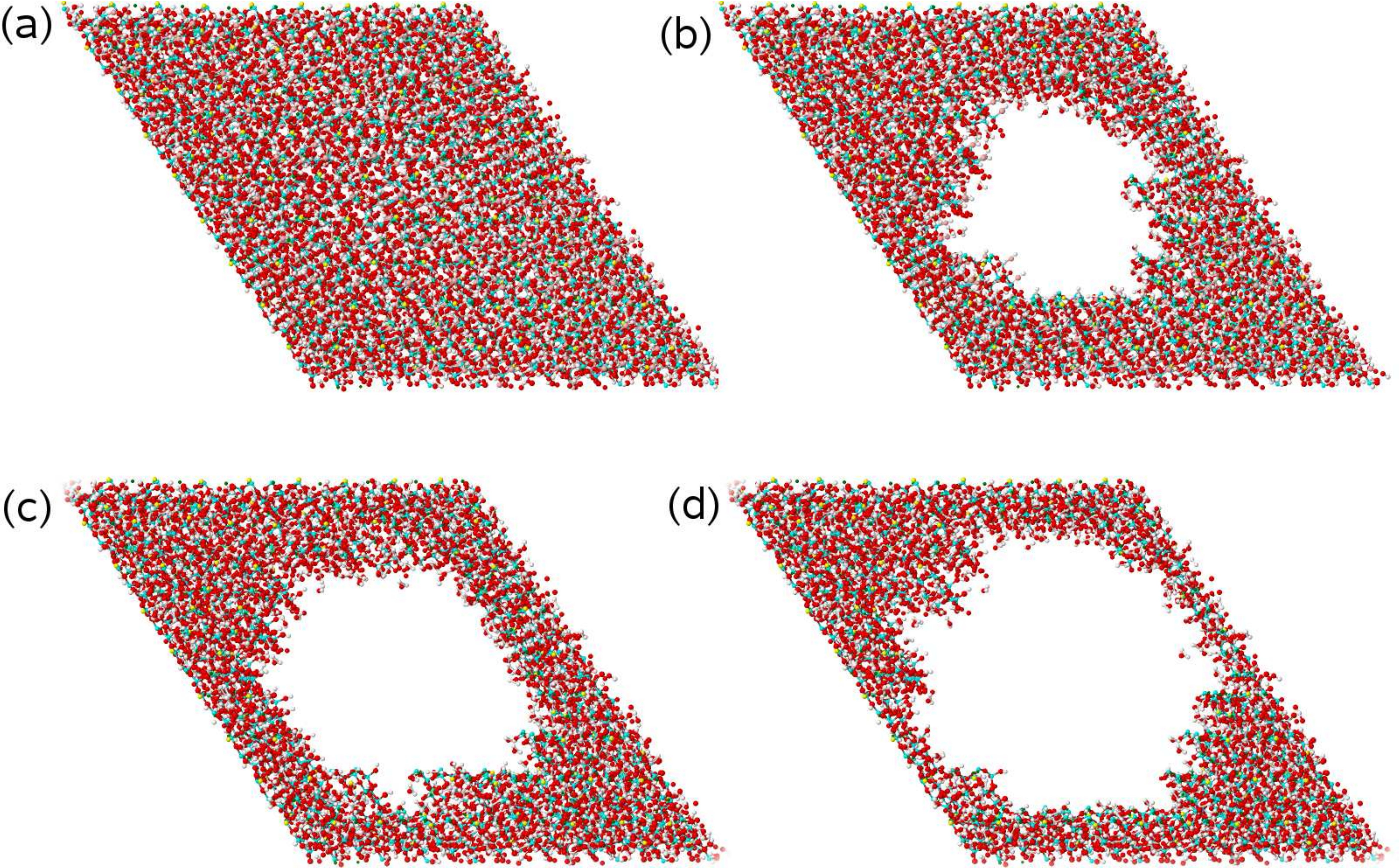}
\caption{Atomistic view of $9\times9\times2$ supercell for tobermorite 14 \AA~with porosities of (a) 0, (b) 24.7\%, (c) 35.8\% and (d) 50.0\%.}
\label{fig2}
\end{figure}

To make modeling of porous tobermorite 14 \AA, the monoclinic $9\times9\times2$ supercell with dimensions 61.020 $\times$ 65.835 $\times$ 56.968 \AA~was created, and then, the spherical voids with diameters of 25 \AA, 32.7 \AA~and 40 \AA~were generated by removing all the atoms in the spherical regions.
In this process, if a certain atom of a molecule was included in the void region, all other atoms in the molecule were also removed.
Then the dimensions of voids were slightly enlarged and their shapes became tangled as shown in Fig.~\ref{fig2}.
By counting the number of removed atoms, the porosities corresponding to these porous models were estimated to be 24.7, 35.8 and 50.0\% respectively.
According to the classification of pores~\cite{Mehta2014}, these can be thought as gel pores and even capillary pores.
It is worth noting that according to the previous studies for porous materials the porosity and temperature are the major factors influencing on their thermal conductivities, whereas shape and arrangement of pores are not significant factors~\cite{Fang,Lukes,Coquil}.
Therefore, we expect that the porous models with the tangled spherical voids can be used safely to study thermal conductivity of porous cement paste.
\begin{figure}[!b]
\centering
\includegraphics[clip=true,scale=0.15]{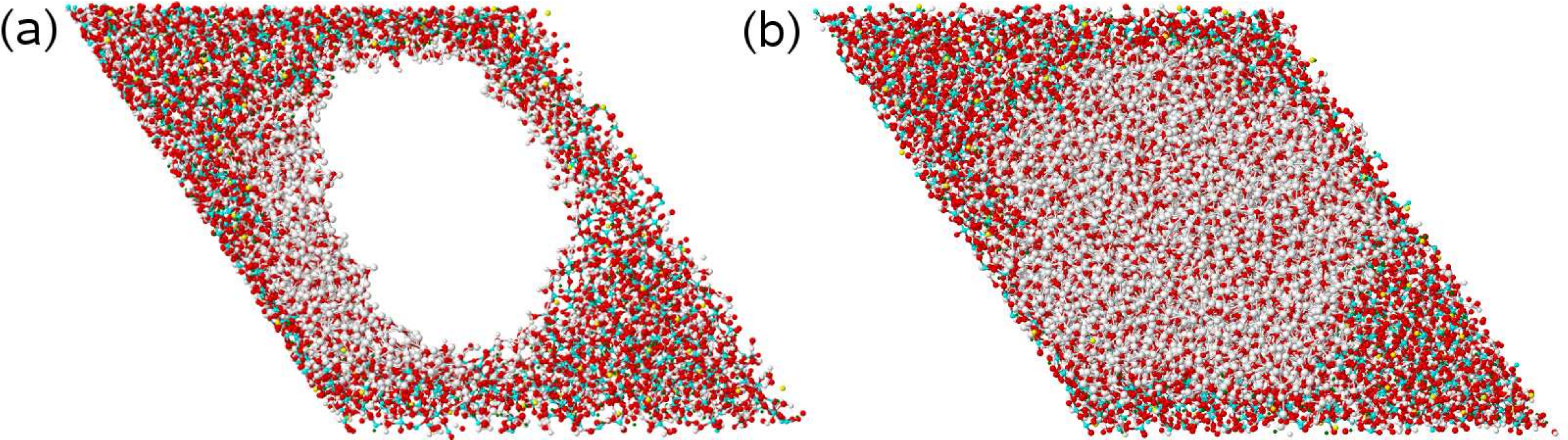}
\caption{Atomistic view of $9\times9\times2$ supercell for porous and moist tobermorite 14 \AA~with (a) 500 and (b) 3187 water molecules in the case of 50\% porosity.}
\label{fig3} 
\end{figure}
We went further into the porous and moist models, which were made by inserting a certain number of water molecules into the pores.
In fact, when the pores inside cement paste and concrete are vacuum or filled with air at dry condition, the concrete wall as a whole can insulate the building effectively against heat conduction due to low thermal conductivity of air (0.026 W/m$\cdot$K~\cite{Lide}).
In reality, however, the inside pores of cement paste and concrete should be filled with water, which can occur during hydration and age or operation of buildings, resulting in dramatic increase of thermal conductivity of concrete wall due to relatively high thermal conductivity of water (0.606 W/m$\cdot$K~\cite{Lide}).
We gradually increased the number of confined water molecules inside the pores, and then, the maximum numbers, which were from equilibrated big water simulation box at 300 K, were determined to be 1574, 2282 and 3187 for the pores with porosities of 24.7, 35.8 and 50.0\% respectively.
Figure~\ref{fig3}) shows the models with 500 and 3187 water molecules in the case of 50\% porosity.

It is necessary to define the moisture content of moist samples.
The moisture content can be defined by mass as~\cite{QingJin},
\begin{equation}
\label{eq_moisture_content_mass}
c_m=\frac{m_w}{m_d},
\end{equation}
where the $m_w$ and $m_d$ are the masses of moist and dry samples respectively.
Also, according to EN ISO 10456~\cite{ISO10456}, we used the definition of moisture content by volume as~\cite{Belkharchouche, Gomes},
\begin{equation}
\label{eq_moisture_content_vol}
c_v=\frac{m_w-m_d}{\rho_{\text{wat}}}\frac{1}{V},
\end{equation}
where $\rho_{\text{wat}}$ is the water density, being 1000 kg/m$^3$, and $V$ is the volume of the sample.


\subsection{\label{sub_force}Interatomic force field}
In the classical MD simulations, it is critical for accuracy to choose the proper interatomic force field for the subject under study.
There have been developed lots of force field (FF) potentials for cementitious materials, including ClayFF, IFF (interface force field), CementFF, ReaxFF and CSH-FF~\cite{Cygan, Heinz, Freeman, Duin,Shahsavari, Mishra, Lau}.
In this work, we choose the ClayFF potential, which has been the most widely used to study cementitious materials over the years, providing sufficiently reliable results in comparison with experiments~\cite{Mutisya,Hou1,Zhang2017,Tavakoli}.

The ClayFF potential is based on the single point charge (SPC) water model to represent the water, hydroxyl and oxygen$-$oxygen interactions.
In this model, the bond stretching energy is represented by the following harmonic potential,
\begin{equation}
\label{eq_bond}
E_{\text{bond stretch}}=\sum_{i\neq j} k_1(r_{ij}-r_0)^2,
\end{equation}
where $r_{ij}$ is the distance between the $i$th and $j$th atoms, and $k_1$ and $r_0$ are the stretching force constant and equilibrium bond length respectively.
The bending energy of bond angles is written as follows,
\begin{equation}
\label{eq_angle}
E_{\text{angle bend}}=\sum_{i\neq j\neq k} k_2(\theta_{ijk}-\theta_0)^2,
\end{equation}
where $\theta_{ijk}$ is the bond angle between the three atoms, and $k_2$ and $\theta_0$ are the bending force constant and equilibrium bond angle respectively.
The ClayFF potential also includes the Coulombic and van der Waals (vdW) dispersion interactions as follows,
\begin{equation}
\label{eq_coulmb}
E_{\text{Coul}}=\frac{e^2}{4\pi\varepsilon_0}\sum_{i\neq j}\frac{q_iq_j}{r_{ij}},
\end{equation}
\begin{equation}
\label{eq_vdw}
E_{\text{vdW}}=\sum_{i\neq j} D^0_{ij} \left[\left(\frac{R^0_{ij}}{r_{ij}}\right)^{12} - 2\left(\frac{R^0_{ij}}{r_{ij}}\right)^6\right],
\end{equation}
where $q_i$ is the partial charge of the $i$th particle, $e$ the elementary charge, and $\varepsilon_0$ the dielectric permittivity of vacuum.
And $D^0_{ij}=\sqrt{D^0_iD^0_j}$ and $R^0_{ij}=(R^0_i+R^0_j)/2$, where $D^0_{i}$ and $R^0_{i}$ are the empirical parameters derived from the experimental data for structural and energetic properties. 
Then, the total energy of the system can be written as the sum of the above force field potentials~\cite{Cygan, Mishra},
\begin{equation}
\label{eq_clayff_total_energy}
E_{\text{total}}=E_{\text{Coul}}+E_{\text{vdW}}+E_{\text{bond stretch}}+E_{\text{angle bend}}.
\end{equation}
We generated the force field parameters for Ca, Si, O and H atoms by using {\it msi2lmp} tool of LAMMPS (Large Scale Atomic/Molecular Massively Parallel Simulator)~\cite{lammps}. 

\subsection{\label{sub_process}Molecular dynamics simulations}
The thermal conductivity $\mathbi{K}$ of material, defined as a second-order tensor coefficient relating the temperature gradient to the heat flux through the Fourier's law, $\mathbi{q}=\mathbi{K}\cdot \nabla T$, can be calculated via MD simulations using two different methods: (1) equilibrium Green-Kubo (GK) method~\cite{Green,Kubo} and (2) non-equilibrium MD (NEMD) or M\"{u}ller-Plathe (MP) method~\cite{Muller}.
In the GK method, which is based on the fluctuation-dissipation theory at an equilibrium state of system, the thermal conductivity  can be calculated by integrating the heat flux autocorrelation function (HFACF) obtained by sufficiently long time run.
Although the thermal conductivity tensor can be obtained through one time simulation, the GK approach has not been applied to the porous materials in the previous works.
Accordingly we also used the GK method for the bulk tobermorite 14 \AA~without any void in this work.

Meanwhile, the NEMD method has been widely used to predict the thermal conductivity of materials with and without porosity, specially porous materials~\cite{Fang,Coquil,Muller,Zhu}.
In this work, we also apply the NEMD method mainly to the porous and moist tobermorite 14 \AA, through at least three simulations to determine the anisotropic thermal conductivity in each direction.
In the NEMD method, the simulation box is divided into an even number of parallel slabs along the measuring direction, and the energy exchange between the 1st and middle slabs is imposed, {\it i.e.}, the velocity of the atom having the lowest kinetic energy of the middle slab is exchanged with the velocity of the atom having the largest kinetic energy of the 1st slab at every proper time step.
At this time, the temperature of the middle slab increases, whereas the temperature of the 1st slab decreases.
Here, the temperature of the $i$th slab is calculated as,
\begin{equation}
\label{eq_slab_temp}
T_i=\frac{1}{3n_ik_B}\sum_{\alpha=1}^{n_i}m_{\alpha}v^2_\alpha,
\end{equation}
where $n_i$ is the number of atoms in the $i$th slab, $k_B$ the Boltzmann constant, and $m_\alpha$ and $v_\alpha$ the mass and velocity of the $\alpha$th atom. 
After sufficiently long simulation time, the temperatures of all the slabs fluctuate around certain values, and the temperature gradient along the slabs should becomes linear.
The periodic boundary condition (PBC) is in general applied in the three crystallographic directions, and then the thermal conductivity is calculated through the Fourier's law as follows,
\begin{equation}
\label{eq_Fourier_law}
K=\frac{J}{2A}\frac{\Delta T}{\Delta x},
\end{equation}
where $J$ is the heat flux that can be obtained from simulation, $\Delta T/\Delta x$ the temperature gradient along the heat flux direction, $A$ the cross sectional area through which the heat flux passes, and the number 2 is due to PBC.

In the NEMD approach, the length of simulation box is a critical factor for accuracy; for the cases of jennite and amorphous silica in the previous works the lengths were set to be larger than 110 \AA~\cite{Coquil} respectively.
In accordance with this, we repeated the $9\times9\times2$ supercell model three times (about 170 \AA) along the [100], [010] and [001] directions respectively for the porous models.
For the case of perfect tobermorite 14 \AA~the supercell was repeated by four times so that the length became 230 \AA.
We set the number of dividing slabs to be 20 in this work.
The energy exchange was conducted at every 100 step during the simulation of $10^6$ steps and further $5\times 10^5$ steps for averaging the temperature difference.

All the simulations in this work were performed by using the LAMMPS code.
The timestep was set to be 0.1 fs, and the cutoff radii for force calculation to be 15 \AA~and 10 \AA~in molecular mechanics (MM) or energy minimization and dynamics respectively.
The $pppm$ solver was used to evaluate the long-range Coulombic interaction.

\section{\label{sec_result}Results and discussion}
\subsection{\label{sub_verify}Assessment of the ClayFF potential for tobermorite 14 \AA}
We assessed the transferability of the ClayFF interatomic potential to tobermorite 14 \AA~through calculation of its lattice parameters, radial distribution functions (RDF) and elastic properties in comparison with the previous experimental and simulation data.
In order to determine the lattice parameters of bulk tobermorite 14 \AA, we conducted the energy minimization of the unit cell by using the conjugate gradient algorithm and the $pppm$ solver, where the lattice constants $a$, $b$, $c$ were controlled independently by external pressure.
As can be seen in Table~\ref{tbl_unit_param}, the lattice parameters were determined to be in good agreement with the experimental values~\cite{Bonaccorsi-tober14} with the maximum relative error of 1.8\%.
\begin{table}[!th]
\small
\caption{\label{tbl_unit_param}The optimized lattice parameters of unit cell of bulk tobermorite 14 \AA~in comparison with previous MD simulations and experimental data.}
\begin{tabular}{l@{\hspace{6pt}}c@{\hspace{6pt}}c@{\hspace{6pt}}c@{\hspace{6pt}}c@{\hspace{6pt}}c@{\hspace{6pt}}c}
\hline
&$a$ (\AA)&$b$ (\AA)&$c$ (\AA)&$\alpha$ ($^\circ$)&$ \beta$ ($^\circ$)&$\gamma$ ($^\circ$)\\
\hline
This work  & 6.780 & 7.315  & 28.484  & 90 & 90 & 123.25 \\
Hajilar {\it et al}.~\cite{Hajilar} & 6.635 & 7.315 & 27.573 & 90 & 90 & 123.25 \\
Mutisya {\it et al}.~\cite{Mutisya} & 6.690 & 7.190 & 27.840 & 90 & 90 & 123.25 \\
Manzano {\it et al}.~\cite{Manzano2009} & 6.760 & 7.340 & 29.150 & 90 & 90.3 & 122.90\\
Exp.~\cite{Bonaccorsi-tober14} & 6.735 & 7.425 & 27.987 & 90 & 90 & 123.25 \\
\hline
\end{tabular}
\end{table}

\begin{figure}[!th]
\centering
\includegraphics[clip=true,scale=0.5]{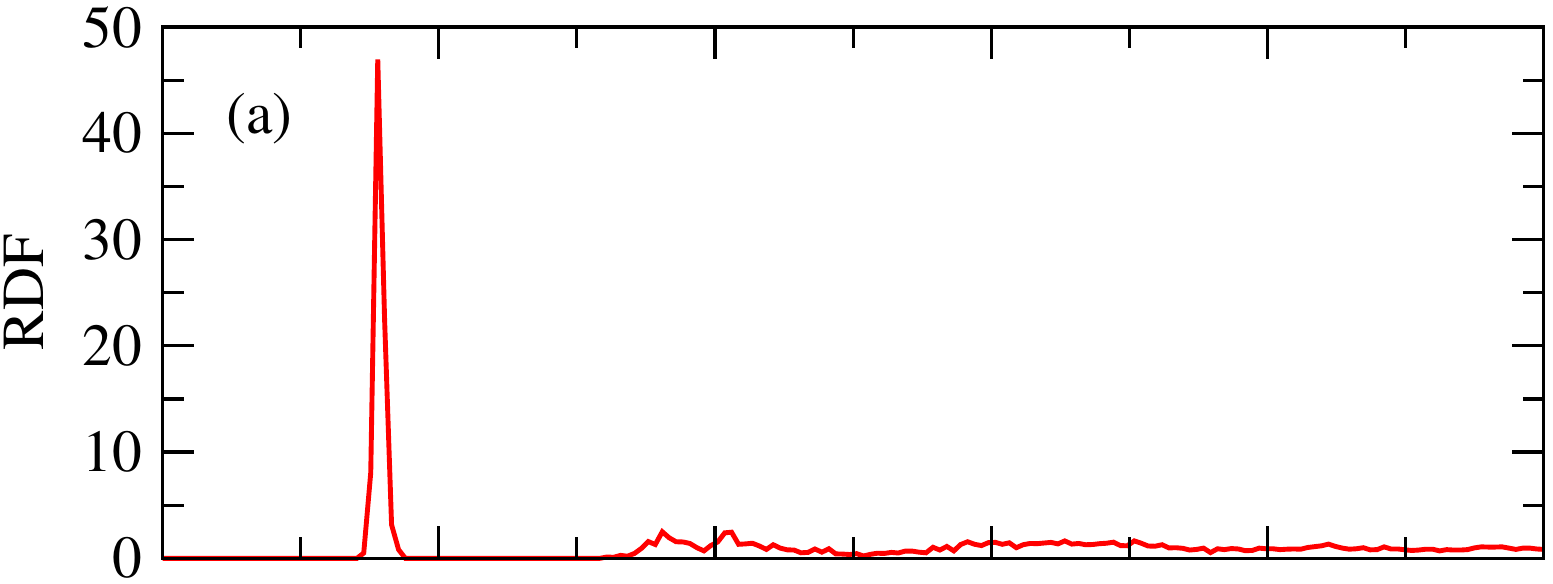} \\
\includegraphics[clip=true,scale=0.5]{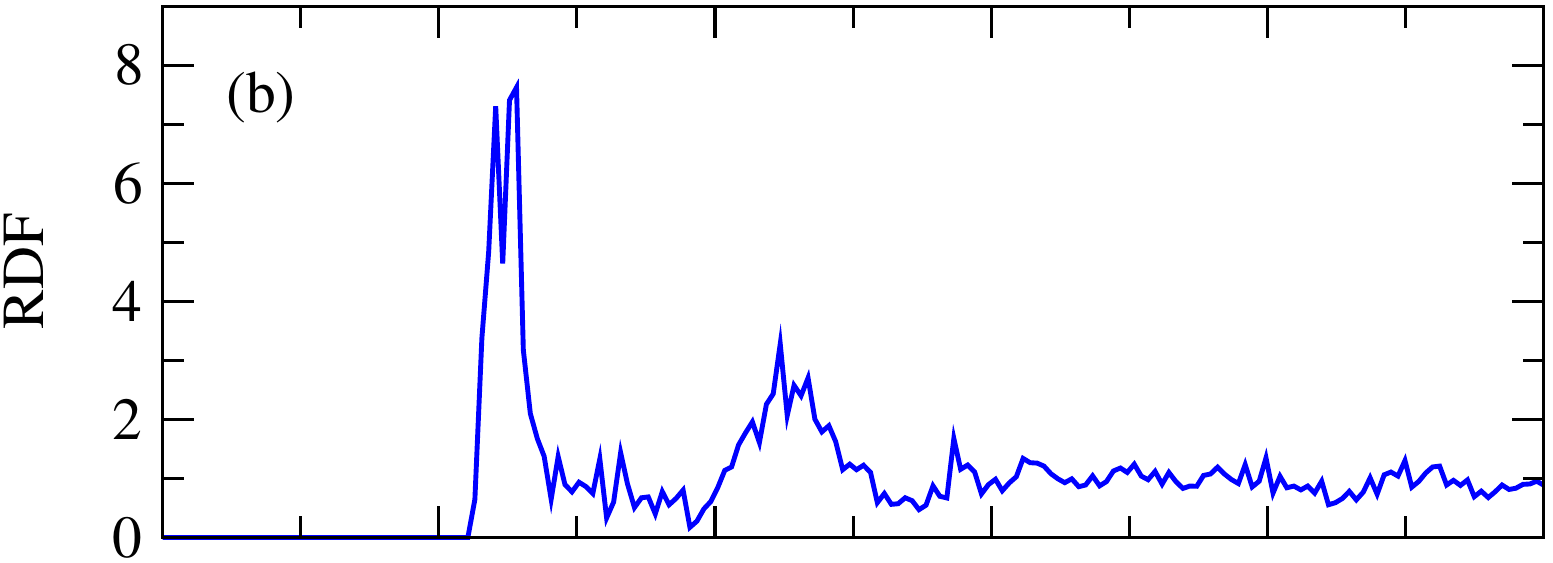} \\
\hspace{4pt}\includegraphics[clip=true,scale=0.5]{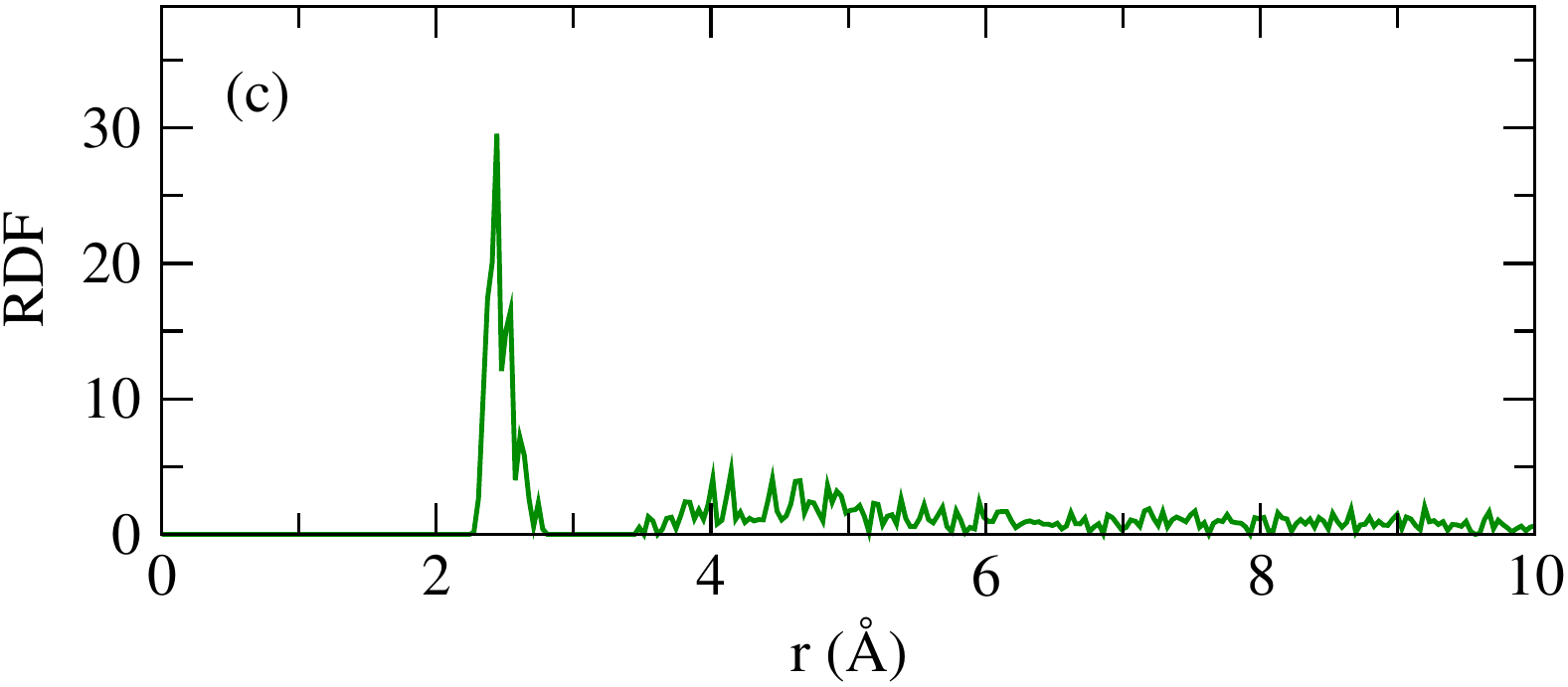}
\caption{Radial distribution function (RDF) for (a) intralayer Si$-$O, (b) intralayer Ca$-$O, and (c) interlayer Ca$-$O in bulk tobermorite 14 \AA, calculated by using the $4\times4\times1$ supercell.}
\label{fig_rdf}
\end{figure}

Figure~\ref{fig_rdf} shows RDFs for Si$-$O and Ca$-$O bonds, calculated by using the $4\times4\times1$ supercell (1664 atoms).
In this figure, one can see that the distance between Si and O atoms is 1.58$\pm$0.04 \AA~and that between the intralayer Ca and O atoms is 2.48$\pm$0.15 \AA, being agreed well with the experimental data of 1.56$\sim$1.74 \AA~and 2.27$\sim$2.74 \AA~\cite{Merlino01,Bonaccorsi-tober14}.
Note that for the case of tobermorite 11 \AA, the intralayer Ca$-$O length is estimated to be shorter as 2.43$\pm$0.10 \AA~\cite{Bowers}.
On the other hand, the distance between the interlayer Ca$_{\text{w}}$ and O$_{\text{w}}$ atoms was calculated to be 2.39$\pm$0.06 \AA~in our simulation (Fig.~\ref{fig_rdf}(c)), which is also in good agreement with the XRD experimental data of 2.36 or 2.55 \AA~\cite{Bonaccorsi-tober14}.

We also calculated the elastic properties of bulk tobermorite 14 \AA~such as bulk modulus $B$, shear modulus $G$, Young's modulus $E$ and Poisson ratio $\nu$.
Table~\ref{tbl_elastic property} presents the calculated values in comparison with the previous classical MM and {\it ab initio} simulations.
Our calculated values are found to be in reasonable agreement with other classical MM simulations using the COMPASS force field conducted by Hajilar {\it et al}.~\cite{Hajilar} and  Al-Ostaz {\it et al}.~\cite{Al-Ostaz} and the Buckingham potential by Manzano {\it et al}.~\cite{Manzano2009} and  {\it ab initio} simulation by Dharmawardhana {\it et al}.~\cite{Dharmawardhana}.
When compared with the available experimental value for the bulk modulus of 47$\pm$3 GPa~\cite{Oh}, our simulated value was slightly overestimated.
These indicate that the ClayFF potential employed in this work
is reasonably accurate and transferable for tobermorite 14 \AA.
\begin{table}[!th]
\small
\caption{\label{tbl_elastic property}The calculated elastic properties including bulk modulus ($B$), shear modulus ($G$), Young's modulus ($E$) and Poisson ratio ($\nu$) of bulk tobermorite 14 \AA~in comparison with the previous classical MM and {\it ab initio} simulations.}
\begin{tabular}{l@{\hspace{6pt}}c@{\hspace{6pt}}c@{\hspace{6pt}}c@{\hspace{6pt}}c}
\hline
& $B$ (GPa) & $G$ (GPa) & $E$ (GPa) & $\nu$ \\
\hline
This work                                         & 54.67& 29.24& 74.45& 0.273\\
Hajilar {\it et al}.~\cite{Hajilar}               & 55.74 & 26.67 & 69.01 & 0.29 \\
Manzano {\it et al}.~\cite{Manzano2009}           & 44.80  & 19.00   & 49.94 & 0.31 \\		
Manzano {\it et al}.~\cite{Manzano2007}           & 46    & 39    & 91    & 0.17 \\
Al-Ostaz {\it et al}.~\cite{Al-Ostaz}             & 52.89 & 18.55 & 49.82 & 0.343\\        
Dharmawardhana {\it et al}.~\cite{Dharmawardhana} & 56.42 & 31.65 & 80.00 & 0.260 \\
Exp.~\cite{Oh} & 47$\pm$3 & $-$ & $-$ & $-$ \\
\hline
\end{tabular}
\end{table}

\subsection{\label{sub_bulk}Thermal conductivity of bulk phase}
\begin{figure}[!b]
\centering
\hspace{4.5pt}\includegraphics[clip=true,scale=0.5]{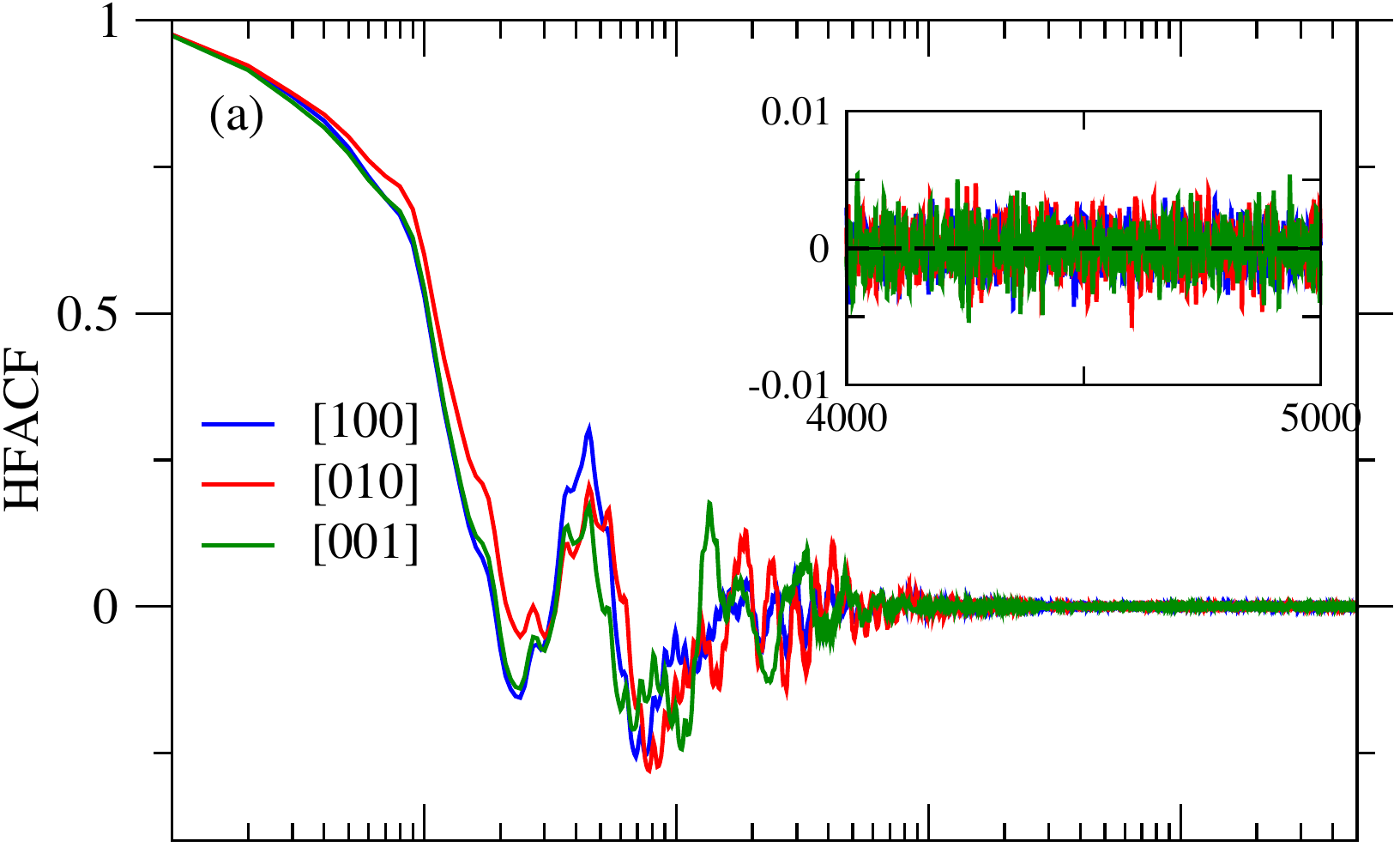}
\includegraphics[clip=true,scale=0.5]{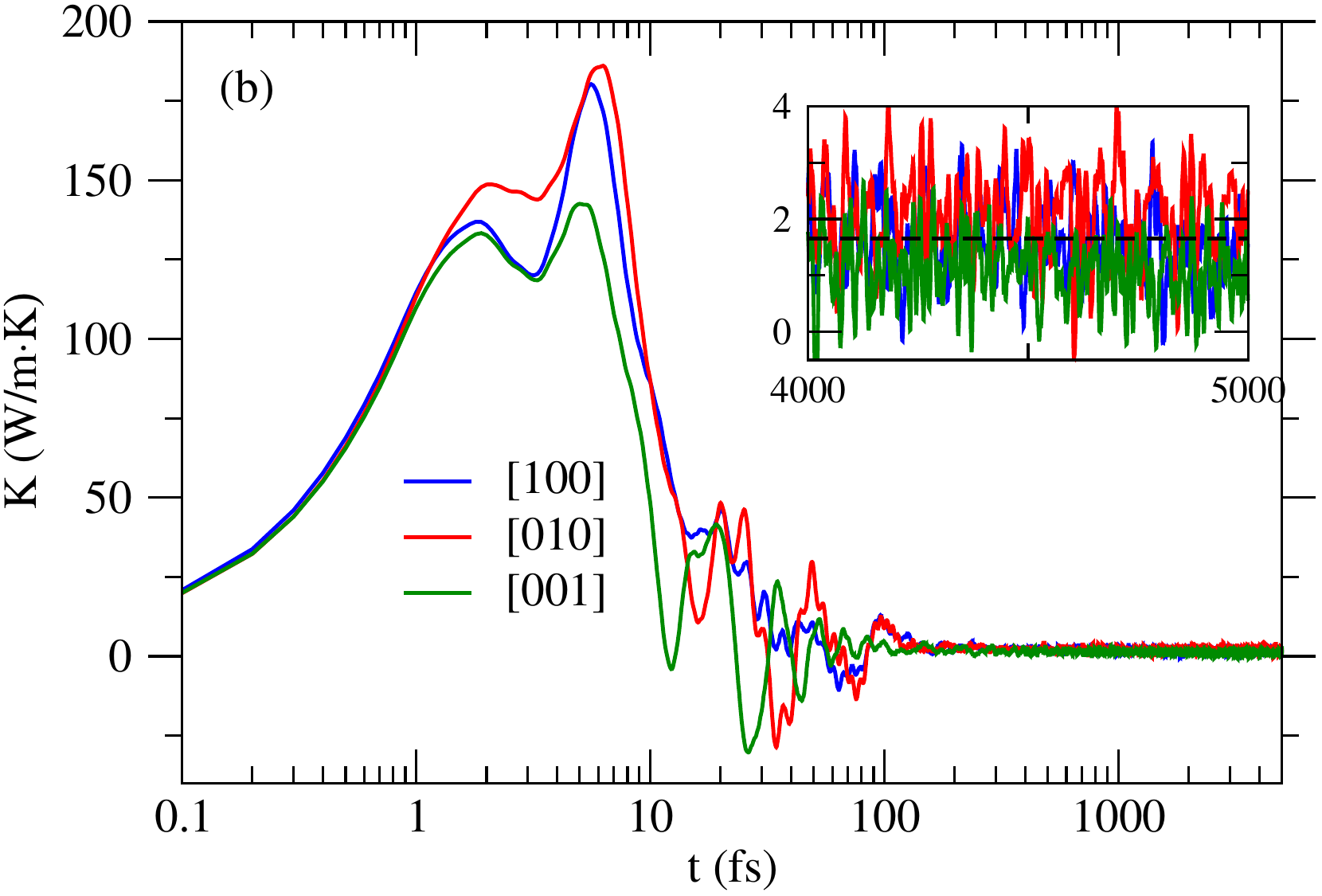}
\caption{(a) Heat flux autocorrelation function (HFACF) and (b) thermal conductivity $K$ of bulk tobermorite 14 \AA~in different crystallographic directions.
Insets show fluctuations of HFACF and the thermal conductivity from the time step of 4000 fs to 5000 fs, where dash lines indicate the average values.}
\label{fig_green_kubo}
\end{figure}
As a preliminary work, we calculated the thermal conductivity of bulk tobermorite 14 \AA~by using both the GK equilibrium and MP NEMD methods.
Comparison with the available previous data was provided.
For applying the GK method, we created the $4\times 4\times 1$ supercell (1664 atoms) as this method is less size dependent, and performed $NVE$ simulation for 10 ns ($10^8$ steps) with a time step of 0.1 fs.
Figure~\ref{fig_green_kubo} shows HFACFs and the thermal conductivities determined from those along the [100], [010] and [001] directions at 300 K.
After the initial transition period of $\sim$200 fs, the system was found to reach the equilibrium state as showing the fluctuation of HFACF around zero point, and the thermal conductivities converged to a certain value.
The principal elements of thermal conductivity tensor were calculated to be $K_1=1.704\pm0.281$ W/m$\cdot$K, $K_2=2.092\pm0.325$ W/m$\cdot$K and $K_3=1.149\pm0.245$ W/m$\cdot$K in the [100], [010] and [001] directions respectively, giving the averaged volumetric value of $K_\text{v}=1.648\pm0.283$ W/m$\cdot$K.
These values are clearly bigger than those of tobermorite 11 \AA~of 0.98$\pm$0.2 W/m$\cdot$K, obtained from CSH FF potential GK simulation by Qomi {\it et al.}~\cite{Qomi1}

As for the anisotropic aspect, the relation of $K_2 > K_1 > K_3$ was found like the jennite model.
The biggest value of $K_2$ is probably due to fairly heat conductive Si$-$O bonds that are aligned along the [010] direction~\cite{Bonaccorsi-jen}, while the smallest value of $K_3$ is due to the interlayer water molecules and looser Ca$-$O bonds, which scatter phonons in the [001] direction~\cite{Qomi1,Bonaccorsi-jen,Dongshuai}.
In contrast to this, Qomi {\it et al.}~\cite{Qomi1} found the different relation of $K_1 > K_2$ due to their different model, which contain many defects in the silicate chains along the [010] direction~\cite{Pellenq}.

For another check, we also performed the NEMD simulations to compute the thermal conductivity of bulk phase, using the simulations boxes obtained by repeating the $9\times9\times2$ supercell model four times in the [100], [010] and [001] directions (about 230 \AA~and 67 392 atoms).
The principal values of thermal conductivity were determined to be $K_1=1.596\pm0.184$, $K_2=2.159\pm 0.25$ and $K_3=0.921\pm 0.232$ W/m$\cdot$K and the averaged value of $K_\text{v}=1.559\pm 0.222$ W/m$\cdot$K.
These are more or less agreed with those by the GK method, and moreover, the same anisotropic feature of $K_2 > K_1 > K_3$ was observed.

\subsection{\label{sub_cement}Thermal conductivity of cement paste}
From the calculated thermal conductivity of bulk tobermorite 14 \AA, the thermal conductivity of cement paste can be estimated by using the upscaling model.
In fact, cement paste is known to be a multiphase porous composite, consisted of C$-$S$-$H ($50\sim 70\%$ of the total volume of cement paste), CH ($20\sim 25\%$), ettringite ($15\sim 20\%$) and hydrogarnet (about 2\%).
Therefore, the thermal conductivity of cement paste should be calculated by averaging the values of these phases.
For instance, Qomi {\it et al.}~\cite{Qomi1} evaluated the mean free path of phonons in various cement clinkers and C$-$S$-$H by applying the kinetic theory of heat transport, finding that these lengths were on the order of Ca$-$O and Si$-$O bond lengths.
Therefore, it can be concluded that the heat transport in C$-$S$-$H and CH is in the vibration-dominated regime and the mean-field homogenization theory can be used in upscaling model for cement paste.

Hence we applied the self-consistent (SC) model derived from the mean-field homogenization theory~\cite{Qomi1} to evaluate the thermal conductivity of porous tobermorite 14 \AA~and cement paste by using the following equation, 
\begin{equation}
\label{eq_sc}
K_{v}^\text{SC}=\frac{\sum_{s=1}^{n_p}{f_sK_v^sB_s^\text{sph}}}{\sum_{s=1}^{n_p}{f_sB_s^\text{sph}}}
\end{equation}
where $n_p$ is the number of phases, and $K_v^s$, $f_s$ and $B_s^\text{sph}=3K_v^\text{SC}/(2K_v^\text{SC}+K_v^s)$ are the volumetric thermal conductivity, the volume fraction and the spherical localization factor of the $s$th phase.
The volumetric thermal conductivity of bulk tobermorite 14 \AA~was treated by using the average of the values obtained by the two different approaches in this work.
Considering the various porosities of low-density (porosity, 36\%) and high-density (porosity, 24\%) and various wet conditions of dry and fully saturation by water, the thermal conductivity of C$-$S$-$H paste was estimated to be 0.766$\sim$1.312 W/m$\cdot$K by using Eq.~\ref{eq_sc}.
Using the previously calculated thermal conductivity of CH as 1.32 W/m$\cdot$K~\cite{Qomi1} and assuming that the cement paste is composed of 70\% C$-$S$-$H and 30\% CH, the thermal conductivity of cement paste was determined to be 0.911$\sim$1.314 W/m$\cdot$K.
Table~\ref{tbl_TC_cement_paste} lists the experimentally determined thermal conductivity of cement paste, indicating that our calculated value is in the range of experimental values~\cite{Yoon,Bentz,Mounanga}. 
\begin{table}[!h]
\small
\caption{\label{tbl_TC_cement_paste}Comparison of the calculated thermal conductivity of cement paste with the available experimental data at various water to cement ratios (w/c).}
\begin{tabular}{lll}
\hline
& $K$ (W/m$\cdot$K)  & w/c\\
\hline
This work  & 0.911 $\sim$ 1.314  & --  \\
Yoon {\it et al}.~\cite{Yoon} 	& 1.28  & 0.4  \\
Xu {\it et al}.~\cite{Xu}     	& 0.53  & 0.35  \\
Bentz~\cite{Bentz} & 0.90 $\sim$ 1.05  & 0.3, 0.4  \\
Hansen {\it et al}.~\cite{Hansen} & 0.88 (early) & 0.5  \\
		& 0.78 (late) &  \\
Mounanga {\it et al}.~\cite{Mounanga} & 1.0 (fresh) & 0.348  \\
		& 1.07 (28 days) &   \\
\hline
\end{tabular} \\
\end{table}

\subsection{\label{sub_moisture}Effect of moisture content on thermal conductivity}
As described above, the porous samples with different porosities of 24.7, 35.8 and 50\% (see Fig.~\ref{fig2}) were created using the $9\times 9\times 2$ supercell, and gradually increasing numbers of water molecules were added into the pores (see Fig.~\ref{fig3}).
\begin{figure}[!b]
\centering
\includegraphics[clip=true,scale=0.27]{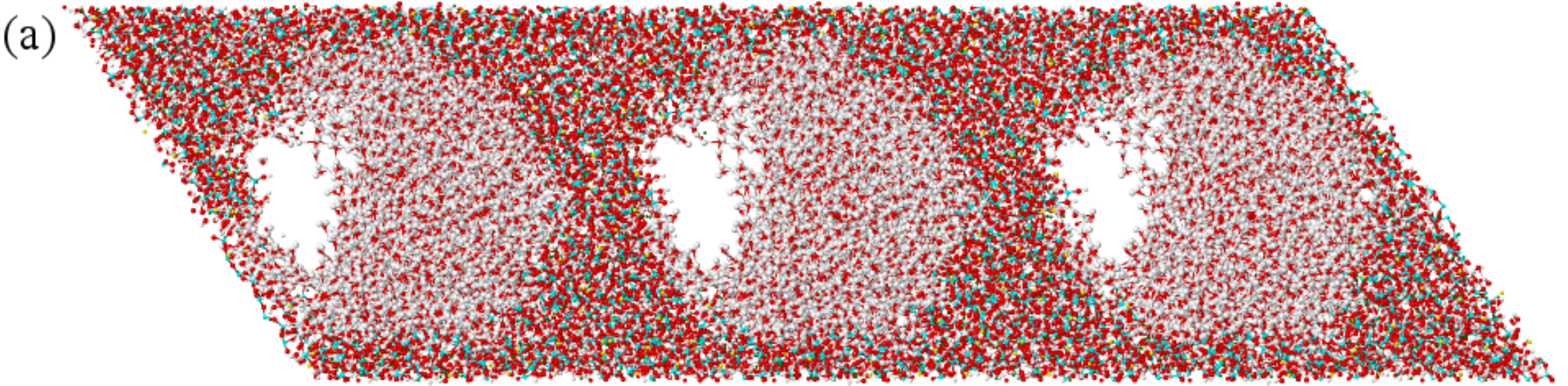}
\includegraphics[clip=true,scale=0.5]{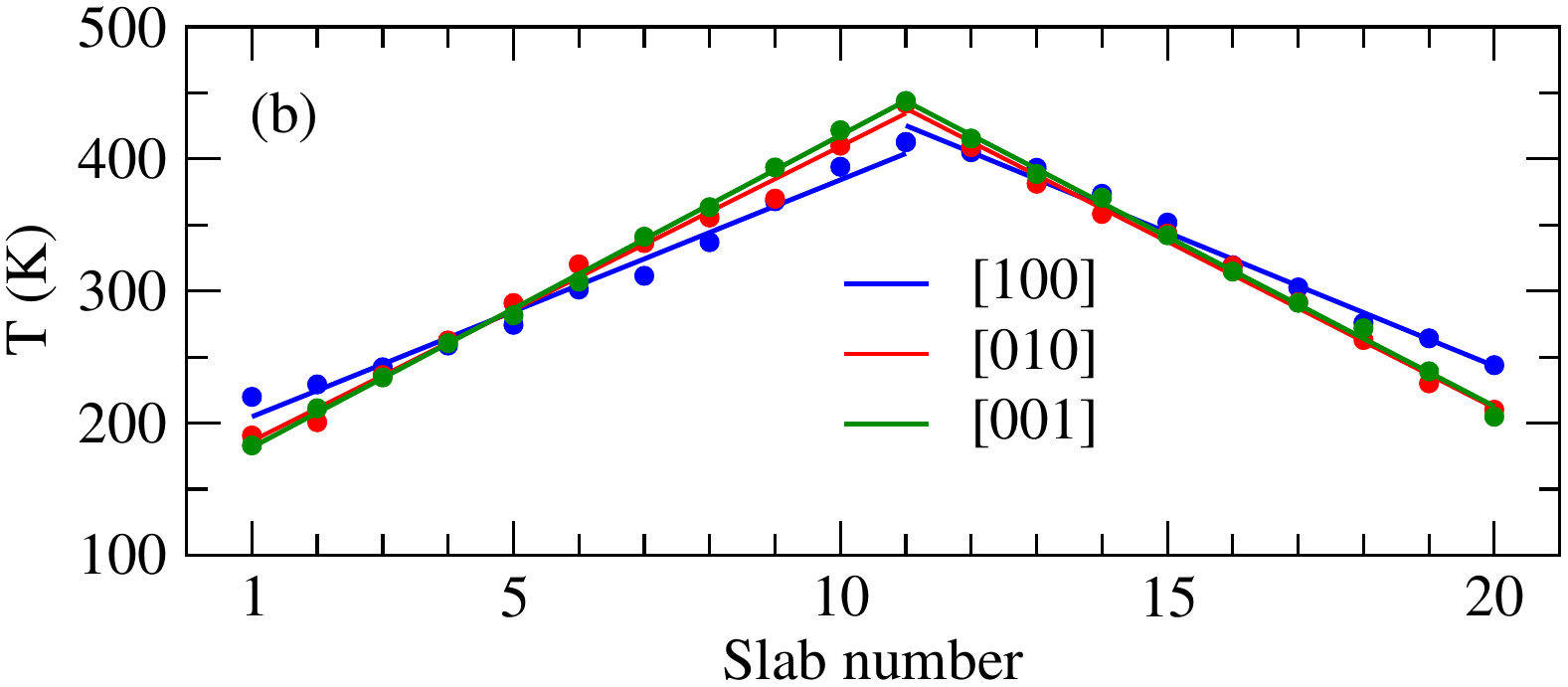}
\caption{(a) Simulation box with porosity of 50\% and 2000 inserted water molecules, being extended in [010] direction and equilibrated through $NVT$ run, and (b) temperature profile along slabs at equilibrium state.}
\label{fig_box}
\end{figure}
These samples with different porosities and water contents were equilibrated through $NVT$ runs at 300 K for $10^6$ steps.
Then the extended samples in each crystallographic direction were generated by repeating the equilibrated samples 3 times along the desired direction, and again equilibrated through $NVT$ runs with $10^6$ steps.
Finally $NVE$ simulations were performed for $10^6$ and $5\times10^5$ steps to measure their thermal conductivities with the MP approach.
The above process was performed 3 times for each direction of [100], [010] and [001].
Figure~\ref{fig_box} typically shows the simulation box extended in [010] direction, with 2000 water molecules and the porosity of 50\%, and the temperature profile along the slabs at the equilibrium state by the kinetic energy exchange.
In the figures, water molecules were observed to be accumulated in the vicinity of pore surface and thus remain voids with small sizes, and the temperature gradient was seen to be nearly linear.

\begin{figure}[!t]
\centering
\includegraphics[clip=true,scale=0.5]{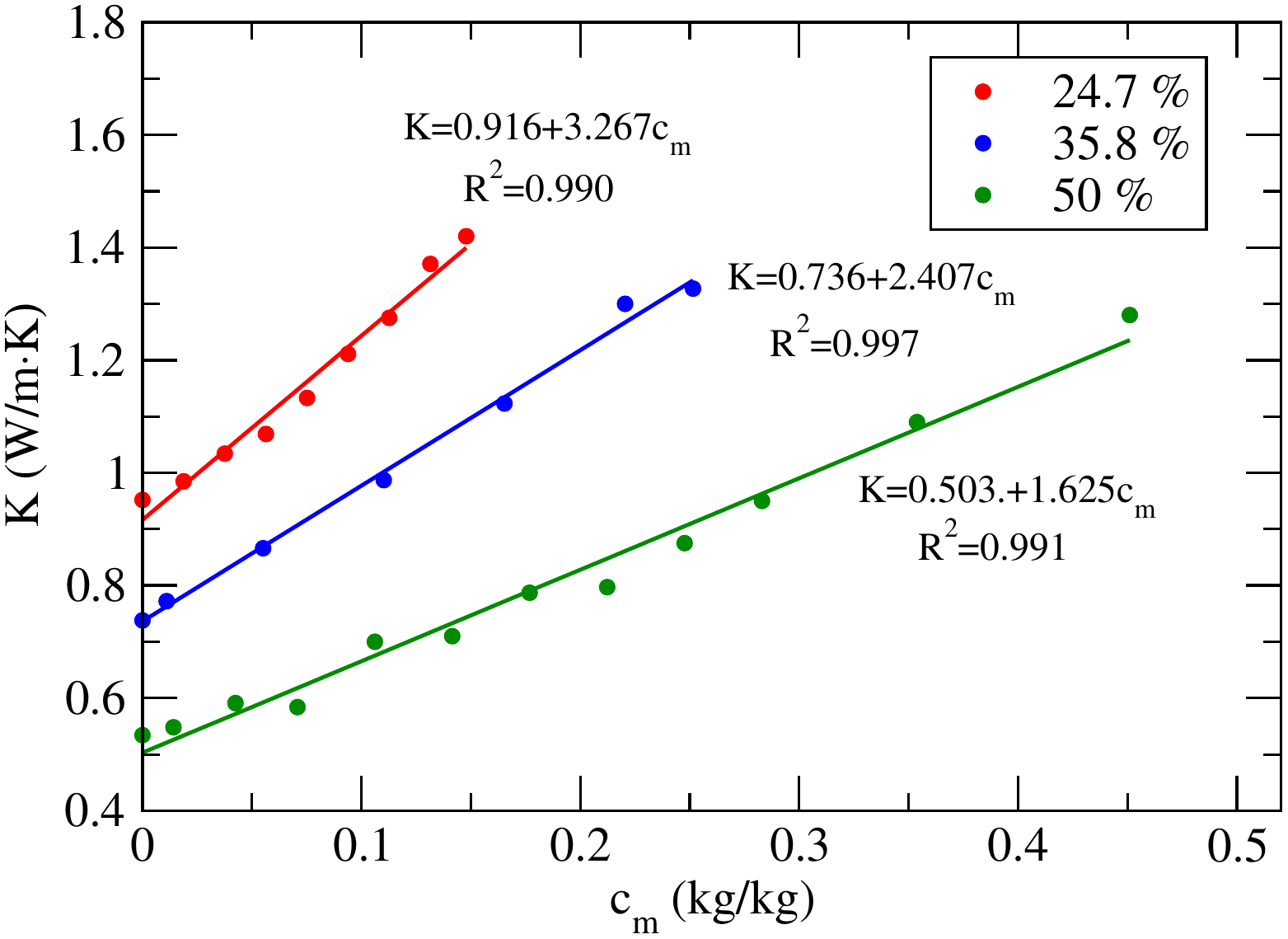}
\caption{Thermal conductivity of porous and moist tobermorite 14 \AA~as a function of moisture content by mass as increasing the porosity from 24.7\% to 35.8\% and to 50\%. Regression lines with slopes and correlation coefficients $R^2$ are presented.}
\label{fig_moist_TC_line}
\end{figure}

The volumetric thermal conductivities of porous and moist samples were obtained by averaging the three different values in the three different directions.
Figure~\ref{fig_moist_TC_line} shows the volumetric thermal conductivity of porous and moist tobermorite 14 \AA~as increasing the porosity and water content.
It was found that the thermal conductivity decreases as increasing the porosity from 24.7\% to 35.8\% and to 50\% while increases as a linear function of moisture content by mass (Eq.~\ref{eq_moisture_content_mass}).
As shown in Fig.~\ref{fig_moist_TC_line}, the linear coefficients derived from regression lines were found to decrease as increasing the porosity in agreement with the experimental observation~\cite{QingJin}.
%

\begin{figure}[!th]
\centering
\includegraphics[clip=true,scale=0.5]{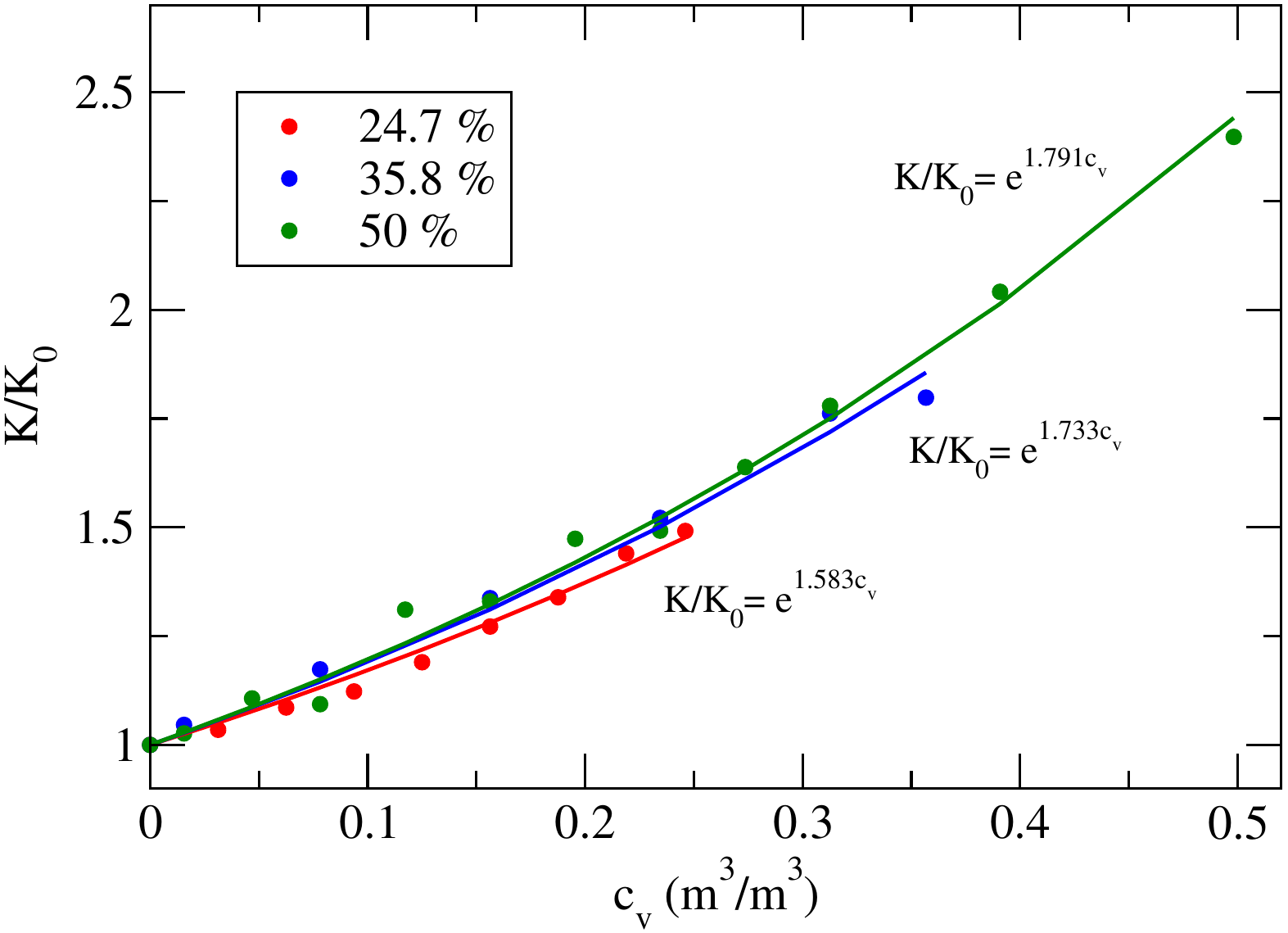}
\caption{Normalized thermal conductivity of porous and moist tobermorite 14 \AA~as an exponential function of moisture content by volume.}
\label{fig_moist_TC_exp}
\end{figure}

In Fig.~\ref{fig_moist_TC_exp}, we show the normalized thermal conductivity obtained versus dry tobermorite 14 \AA~($K_0$) as a function of volume moisture content as suggested by standards~\cite{ISO10456,EN1745}.
Unlike the thermal conductivity, the normalized ones were found to fit into the exponential functions with correlation coefficients of over 0.996.
The exponents in the lines were obtained to be 1.583, 1.733 and 1.791 for the porosities of 24.7, 35.8 and 50\% respectively.
When compared with the experimental result 4.24 by Gomes {\it et al.}~\cite{Gomes} and the standard value 4.00 for mortar with mass densities of 250$\sim$2000 kg/m$^3$~\cite{ISO10456,EN1745}, our simulated exponents are clearly smaller.
This is probably due to the different density (or porosity) and different components of mortar; in our case only tobermorite 14 \AA~as C$-$S$-$H paste was considered but other components of mortar such as sand and CH were not included.
The density and porosity are also different; in the experimental works the densities of cement concretes were 415$\sim$630 kg/m$^3$ (porosities of 76$\sim$84\%)~\cite{QingJin} and 430$\sim$830 kg/m$^3$~\cite{Gomes}, which are lighter than our density values of 1104$\sim$1662 kg/m$^3$ (porosities of 24.7$\sim$50 \%).
Here it should be emphasized that the thermal conductivity of porous and moist tobermorite 14 \AA~increases exponentially when the moisture content increases and the exponent increases with the increase of porosity.

\begin{figure}[!th]
\centering
\includegraphics[clip=true,scale=0.5]{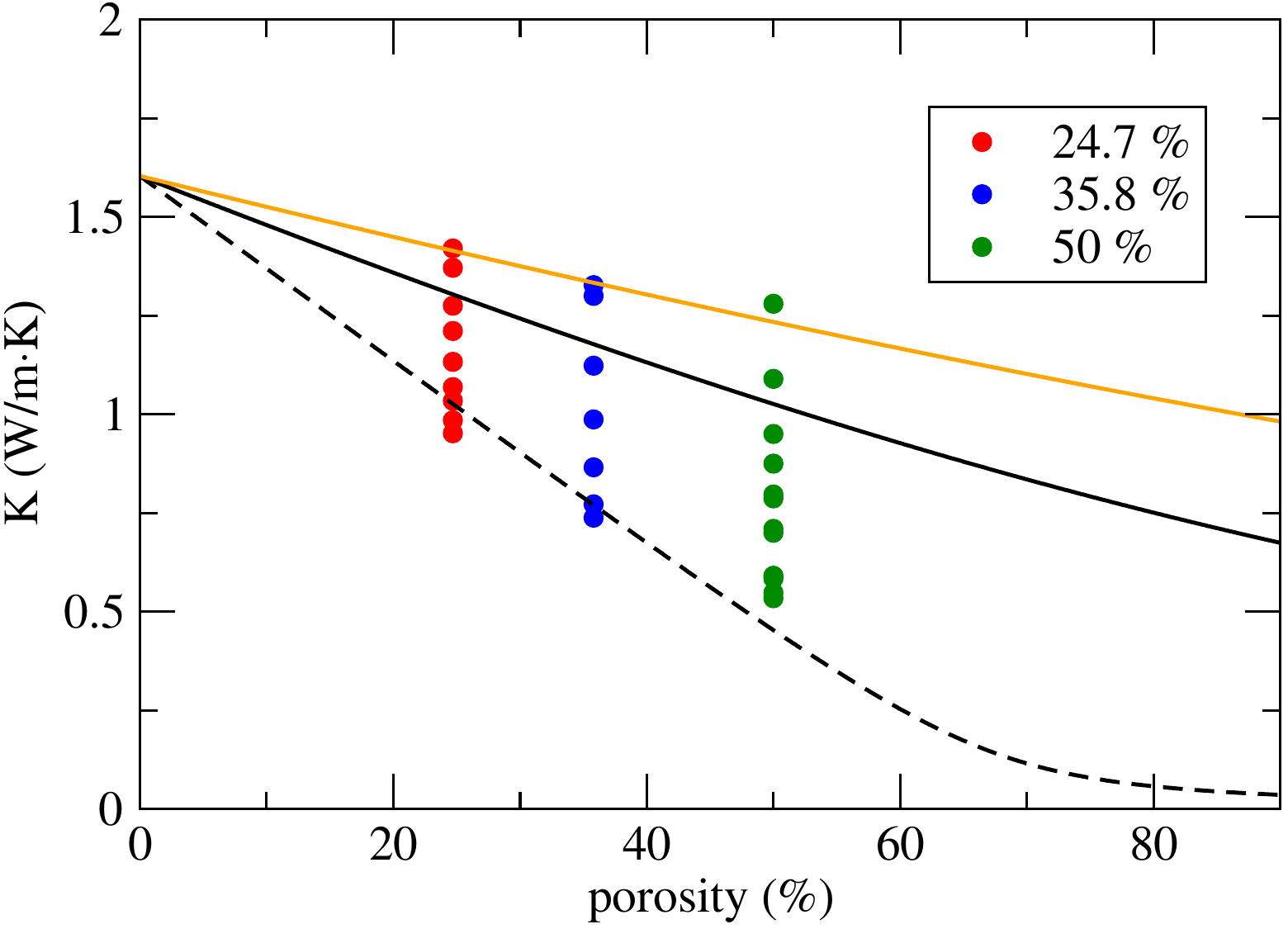}		
\caption{Thermal conductivities of porous and moist tobermorite 14 \AA~as a function of porosity as increasing the water content. Black solid and dash lines are obtained by self-consistent (SC) model at fully wet and dry conditions respectively, and orange line indicates those of confined water calculated by the GK approach.}
\label{fig_porosity_moist}
\end{figure}

We also analyzed the data as an explicit function of porosity. Figure~\ref{fig_porosity_moist} shows the thermal conductivities as a function of porosity, together with the analysis lines predicted by the SC model for dry and fully wet porous tobermorite 14 \AA.
At a given porosity the thermal conductivities should lay in the vertical line to the porosity axis between the lines.
In this figure, one can see that the thermal conductivities of dry porous tobermorite 14 \AA~decreases properly according to the SC model line ({\it i.e.}, the minimum values at the porosities of 24.7, 35.8 and 50\% are seen nearby the dash line).
When the pores were fully saturated by water molecules, however, the simulated thermal conductivities were found to be located over the SC model line ({\it i.e.}, the maximum values are observed above the black solid line), probably due to the confined water molecules in the pores.
Although the thermal conductivity of bulk water was reported to be 0.6062 W/m$\cdot$K~\cite{Lide}, those of confined water in the nano-size pore should be remarkably changed as their atomistic structure and thermodynamics vary in the C$-$S$-$H gel~\cite{Qomi1,Hou2,Bordallo,Goracci}.
In fact, the trapped water within the C$-$S$-$H gel was reported to exhibit the features of supercooled water~\cite{Qomi1} and the diffusion coefficient of the pore water was found to significantly reduce due to that the stable hydrogen bonds connected with Ca$-$OH and Si$-$OH groups restrict the transport of the surface water molecules~\cite{Hou2}.
At the position of 10$\sim$15 \AA~away from the surface of pore, however, the confined water molecules can be expected to behave bulk water.
According to the quasielastic neutron scattering experiment~\cite{Bordallo}, the confined water in C$-$S$-$H can be distinguished into chemically bound water, physically bound water and unbound water.
The chemically or physically bound water was regarded to interact with the surface atom of the gel pore, whereas the unbound water to be confined within the capillary pore~\cite{Bordallo}.
The diffusion coefficients of gel pore water were measured to be $D\approx 10^{-10}$ m$^2$/s and those of capillary pore water to be $D\approx 10^{-9}$ m$^2$/s~\cite{Bordallo}, which are significantly smaller than those of bulk water, due to the interaction of the water molecules with C$-$S$-$H.
Based on this consideration, we can expect that the thermal conductivity of confined nano-pore water in C$-$S$-$H is larger than that of bulk water.
%

To prove this expectation, we calculated the thermal conductivity of water confined in the nano-pore with a diameter of 25 \AA~(porosity 24.7\%) by applying the GK approach.
The thermal conductivity of confined water converged to 0.9257 W/m$\cdot$K, and by using this value and applying the SC model, the analysis line was obtained as shown by orange line in Fig.~\ref{fig_porosity_moist}.
Then the maximum thermal conductivities of porous and moist tobermorite 14 \AA~lay on the orange line.
As increasing the diameter of pore, the thermal conductivity of confined water will decrease, approaching to the value of bulk water.

We developed a new kind of multifunctional composite building materials that has functions of good thermal insulation and moisture proofing for concrete buildings~\cite{Ri18wipo}.
As possible mechanism for such multifunctions, two-step hydration of cement paste and mineral powder (SiO$_2$ and Al$_2$O$_3$) with particle sizes of 0.5$\sim$50 $\mu$m has been proposed to occur, forming the ceramics layer (porosity 10\%) on the surface of concrete wall~\cite{yucj19mcp}.
This ceramic layer can prohibit the percolation of water molecules from outside environment, which inhibits the concrete wall effectively from increasing the thermal conductivity and facilitate the thermal performance of concrete wall.

\subsection{\label{sub_phonon}Phonon analysis of tobermorite 14 \AA}
We calculated the phonon densities of states (PDOS) of various tobermorite 14 \AA~models since PDOS is related to variety of thermodynamic properties such as thermal conductivity and specific heat capacity.
In general, heat conduction in materials can occur by different heat carriers, which include electrons, phonons and photons.
Among such heat carriers, the most contributer can be different according to the material and temperature.
For dielectric solids, the major heat conduction is through the phonon, of which vibration is harmonic at low temperature.
The PDOS can be calculated with two different methods: the eigenvalue decomposition of the dynamic matrix and the Fourier transform method of the velocity autocorrelation function (VACF)~\cite{Qomi1,Mutisya}.
In this work we used the second method via Fourier transform of VACF, in which the PDOS $g(\omega)$ can be computed by the following equation,
\begin{equation}
\label{eq_PDOS}
g(\omega)=\frac{1}{Nk_BT} \sum_{\alpha=1}^{N} m_\alpha \int_{-\infty}^{\infty}\left\langle v_{\alpha}(t)\cdot v_{\alpha}(0)\right\rangle e^{i\omega t}dt, 
\end{equation}
where $\omega$ is the frequency, $v_\alpha$(t) the velocity of the $\alpha$th atom, and $\langle v_\alpha(t)\cdot v_\alpha(0) \rangle / \langle v_\alpha(0)\cdot v_\alpha(0) \rangle$ is the VACF.

\begin{figure*}[!th]
\centering
\hspace{1pt}\includegraphics[clip=true,scale=0.5]{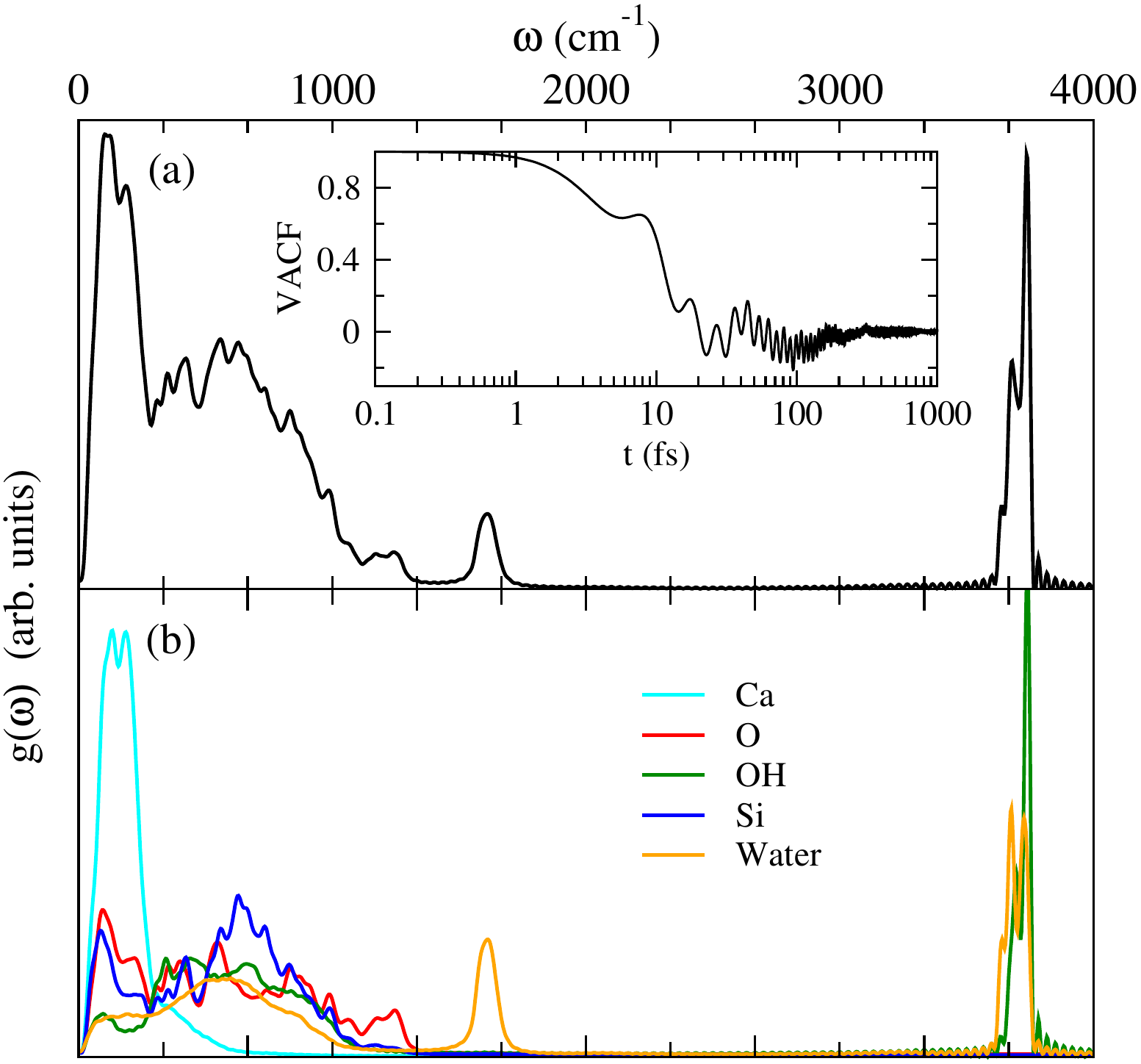}
\includegraphics[clip=true,scale=0.5]{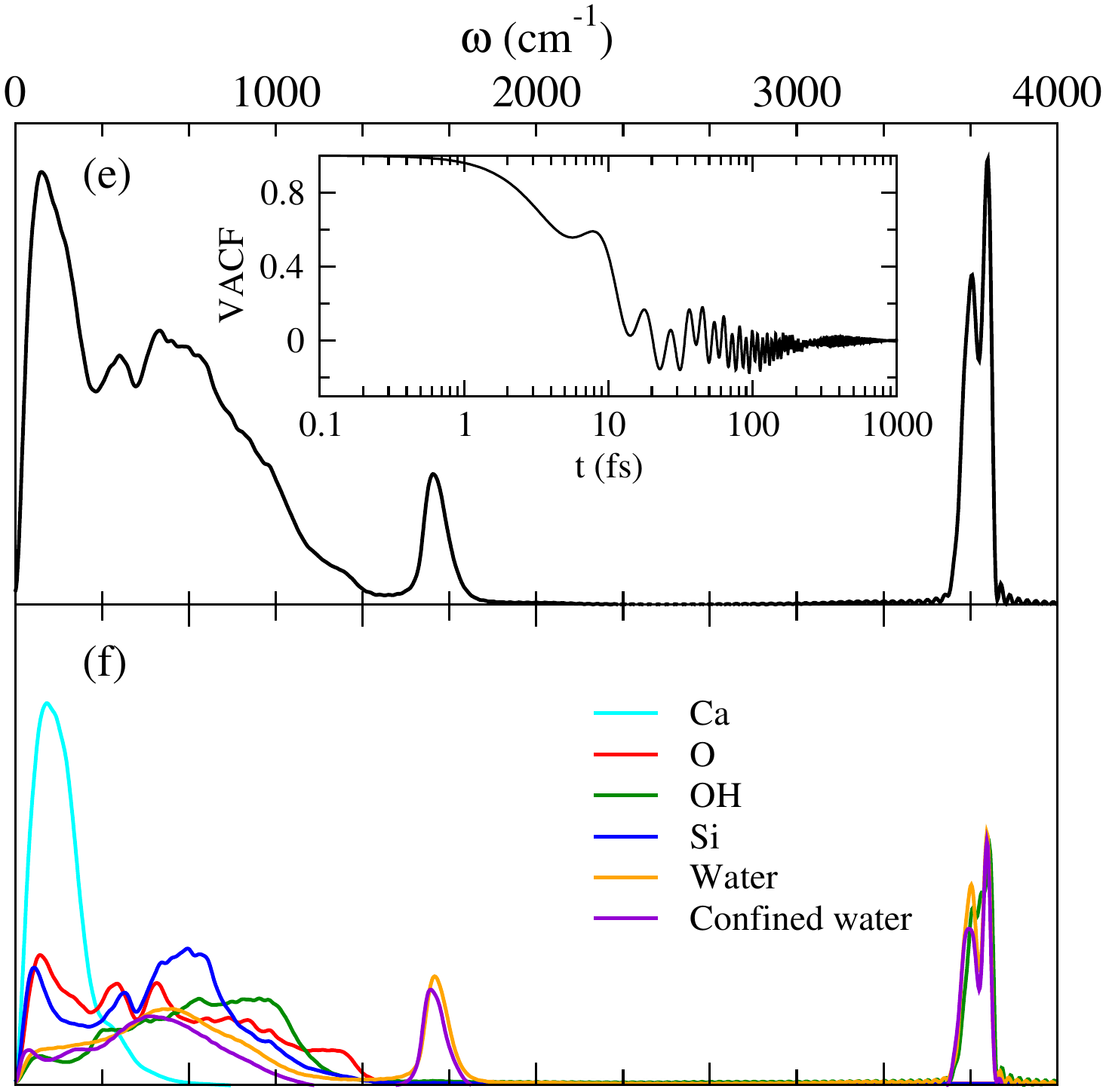} \\
\includegraphics[clip=true,scale=0.5]{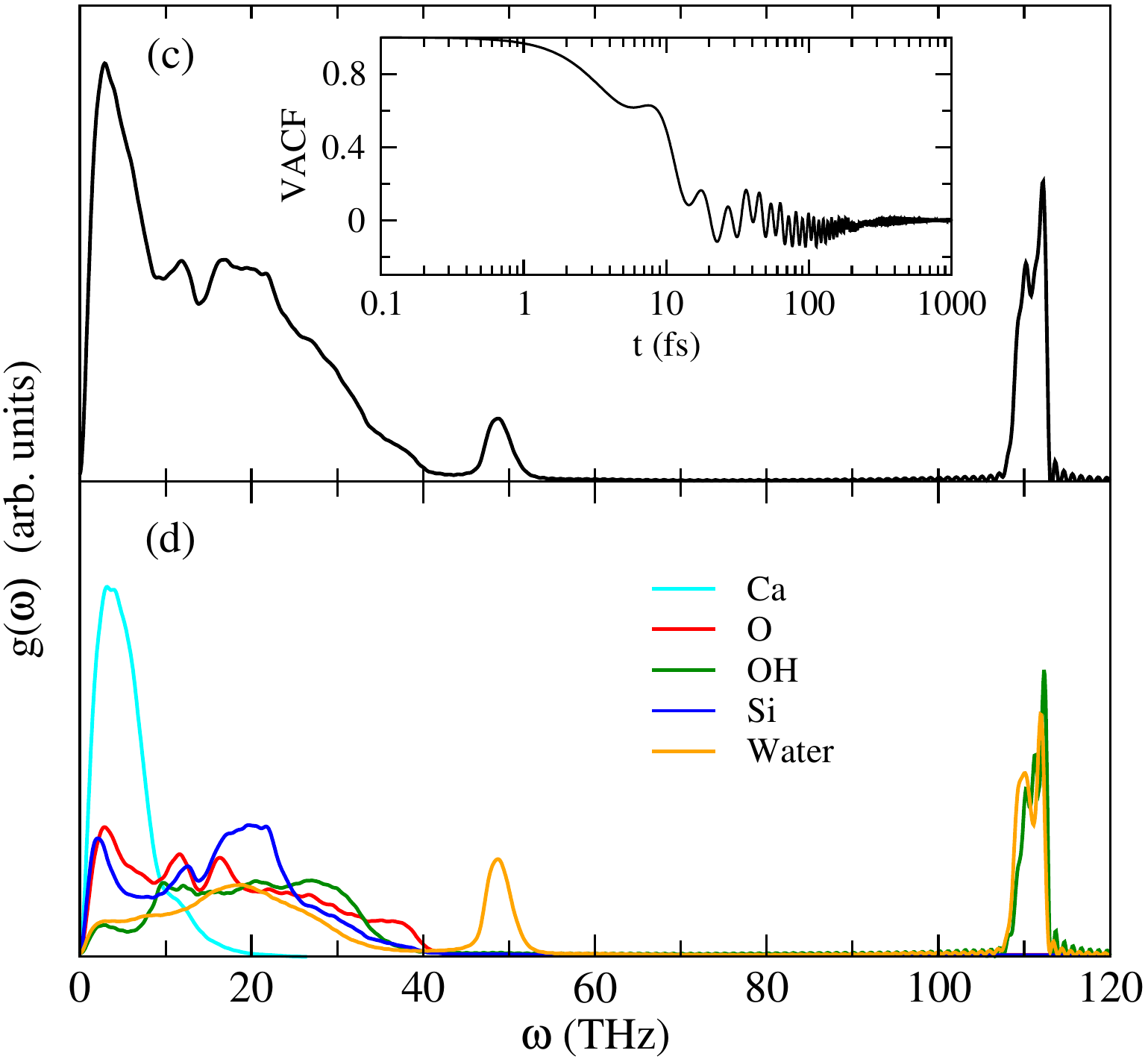}
~\includegraphics[clip=true,scale=0.5]{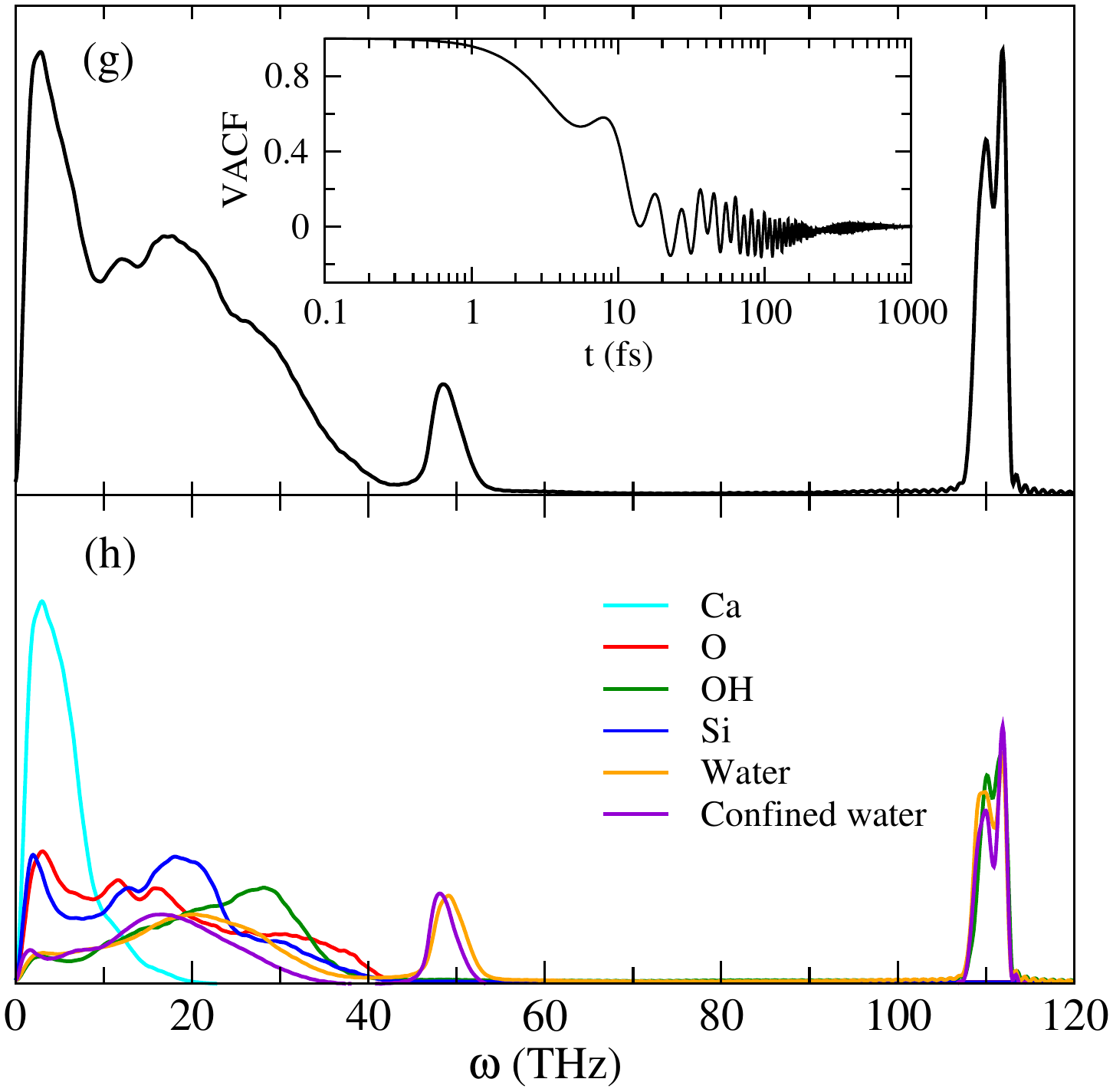}	
\caption{Total (top) and partial (bottom) phonon density of states $g(\omega)$ in (a, b) bulk tobermorite 14 \AA, (c, d) porous sample with a porosity of 50\%, (e, f) porous and moist sample with 500 water molecules, and (g, h) porous and moist sample with 1000 water molecules. Insets show the velocity autocorrelation function (VACF).}
\label{fig_phonon}
\end{figure*}
Figure~\ref{fig_phonon} shows the PDOS of bulk tobermorite 14 \AA, porous sample, and porous and moist sample with 500 and 1000 water molecules in the case of 50\% porosity.
In this figure, the VACFs were observed to fluctuate around zero after about 300 fs of simulation time.
To clarify the individual contributions from Ca, Si, O atoms, OH group, interlayer water and confined water molecules, their partial PDOSs were also analyzed.

In all the partial PDOSs, the range of Ca$-$O bond vibration was found to be $0\sim 430$ cm$^{-1}$, which is in good agreement with the experimental data of $0\sim 400$ cm$^{-1}$~\cite{Yu99phonon} and the simulation data of $0\sim 450$ cm$^{-1}$~\cite{Qomi1,Mutisya,Vidmer14ccr}.
Meanwhile, the range of Si$-$O bond vibration was found to be $0\sim 1200$ cm$^{-1}$ in good agreement with the experiment~\cite{Yu99phonon} and the previous simulations~\cite{Qomi1,Mutisya}.
Here the phonon vibrations with $0\sim 400$ cm$^{-1}$ were found to be associated with the coupled rotation of \ce{SiO4}~\cite{Qomi1}.
One can see the peaks around 420 cm$^{-1}$ and 630 cm$^{-1}$, which are close to the experimentally observed peaks around $400\sim500$ cm$^{-1}$ and 660 cm$^{-1}$, being associated with the deformation of \ce{SiO4} and bending of Si$-$O$-$Si bond angles~\cite{Yu99phonon}.
The phonon vibration with $800\sim 1200$ cm$^{-1}$ was reported to be associated with symmetric and asymmetric stretching of Si$-$O bonds.
For the OH partial PDOS, the phonon of OH species was found to be smaller than that of Si atoms, which is similar to the previous {\it ab initio} work~\cite{Mutisya} but slightly different from the MM work~\cite{Qomi1} due to different C$-$S$-$H models.
The phonon vibrations with 1610 cm$^{-1}$ were found to be associated with bending of the H$-$O$-$H angle of water molecule, being agreed well with the previous works~\cite{Qomi1,Mutisya,Yu99phonon}.
The phonon frequencies of confined water species were found around 1595 cm$^{-1}$.
On the other hand, the phonon vibrations found over 3600 cm$^{-1}$ correspond to the symmetric and asymmetric stretching of O$-$H bonds of water molecules and the stretching of hydroxyl species in the intralayer.
In the porous and moist samples, the peaks of interlayer water, confined water molecules and hydroxyl groups were found at 3660, 3730 and 3740 cm$^{-1}$ respectively.

\section{\label{sec_con}Conclusions}
In this work, we have investigated the thermal conductivity of porous and moist tobermorite 14 \AA~as varying the porosity and the moisture content by using MD simulations with ClayFF interatomic potential.
In the calculation of thermal conductivity, we have applied the equilibrium Green-Kubo and non-equilibrium M\"{u}ller-Plathe methods.
The porous tobermorite 14 \AA~models with various porosities of 24.7, 35.8 and 50\% have been created by removing atoms within the spherical region with the corresponding radii in the $9\times9\times2$ supercell.
Different numbers of water molecules were inserted into the pores so as to make the moist models with different moisture contents.
Through the calculation of structural, elastic and thermal properties, the transferability of ClayFF potential to tobermorite 14 \AA~has been checked.
We determined the thermal conductivity of cement paste by using the self-consistent approach and that of bulk tobermorite 14 \AA~to be 0.911$\sim$1.314 W/m$\cdot$K in good agreement with the experimental data.
We have found that the thermal conductivity of porous and moist tobermorite 14 \AA~increases as a linear function of moisture content by mass, and the slope decreases as the porosity increases, being agreed well with the previous experiment.
On the other hand, the normalized thermal conductivity was found to exponentially increase as the moisture content by volume increases, and the exponent increases as increasing the porosity.
It has been revealed that the calculated thermal conductivities at different porosities and moisture contents are found between the lower and upper limits estimated by self-consistent model using the thermal conductivity of confined water.
Finally the phonon density of states have been calculated through Fourier transformation of velocity autocorrelation function, and analyzed in detail to clarify the contribution of individual atoms to the thermal properties.

\section*{\label{ack}Acknowledgments}
This work has been supported under the state research project ``Design of Innovative Functional Materials for Energy and Environmental Application'' (No. 2016-20) by the State Commission of Science and Technology, DPR Korea.
The simulations have been performed on the HP Blade System C7000 (HP BL460c) that is managed by Faculty of Materials Science, Kim Il Sung University.

\section*{\label{note}Notes}
The authors declare no competing financial interest.

\bibliographystyle{elsarticle-num-names}
\bibliography{Reference}

\end{document}